\documentclass[twocolumn,showpacs,aps,prd]{revtex4}

\usepackage{graphicx}
\usepackage{dcolumn}
\usepackage{amsmath}
\usepackage{epsfig}

\input pubboard/babarsym

\newcommand{\BABARPubYear}    {03}
\newcommand{\BABARPubNumber}  {010}

\newcommand{\SLACPubNumber} {9736}

\def\figurebox#1#2#3{%
    \def\arg{#3}%
    \ifx\arg\empty
    {\hfill\vbox{\hsize#2\hrule\hbox to #2{\vrule\hfill\vbox to #1{\hsize#2\vfill}\vrule}\hrule}\hfill}%
    \else
    {\hfill\epsfbox{#3}\hfill}%
    \fi}

\newcommand{\btoddk}{\ensuremath{B\to \Dbar^{(*)} D^{(*)} K}}
\newcommand{\de}{\ensuremath{\Delta E}}
\newcommand{\dkpi}{\Dz\to\Km\pip}
\newcommand{\dkpipiz}{\Dz\to\Km\pip\piz}
\newcommand{\dkpipipi}{\Dz\to\Km\pip\pim\pip}

\newcommand{\bdzdzk}{$\Bu\to \Dzb\Dz\Kp$}

\newcommand{\btodsdsk}{\ensuremath{\Bu \to \Dstarm \Dstarp \Kp}}

\newcommand{\modevii}{\ensuremath{\Bz\to \Dstarm \Dp \KS}}

\newcommand{\modex}{\ensuremath{\Bz\to \Dstarm \Dstarp \KS}}

\newcommand{\modexvi}{\ensuremath{\Bu\to \Dstarm \Dp \Kp}}

\begin{document}

\preprint{\babar-PUB-\BABARPubYear/\BABARPubNumber}
\preprint{SLAC-PUB-\SLACPubNumber}
\begin{flushleft}
\babar-PUB-\BABARPubYear/\BABARPubNumber\\
SLAC-PUB-\SLACPubNumber\\
\end{flushleft}

\title{
{\large \bf
Measurement of the Branching Fractions for the Exclusive Decays of
\boldmath\Bz and \boldmath\Bu to \boldmath ${\Dbar^{(*)}D^{(*)}K}$}
}

%
\author{B.~Aubert}
\author{R.~Barate}
\author{D.~Boutigny}
\author{J.-M.~Gaillard}
\author{A.~Hicheur}
\author{Y.~Karyotakis}
\author{J.~P.~Lees}
\author{P.~Robbe}
\author{V.~Tisserand}
\author{A.~Zghiche}
\affiliation{Laboratoire de Physique des Particules, F-74941 Annecy-le-Vieux, France }
% ...
\collaboration{The \babar\ Collaboration}
\noaffiliation

\date{\today}

\begin{abstract}
\begin{center}
\large \bf Abstract
\end{center}
We report the observation of  823$\pm$57 $B^0$ and 970$\pm$65 $B^+$  decays 
to doubly charmed final states $\overline D^{(*)}D^{(*)}K$, where $\overline D^{(*)}$ 
and $ D^{(*)}$ are fully reconstructed and $K$ is either a $K^\pm$ or 
a $K^0$.  We use a sample of $82.3\pm 0.9$ million $B \overline B$ events 
collected between 1999 and 2002 with the \babar\ detector at the PEP-II storage
ring at the Stanford Linear Accelerator Center.
The 22 possible $B$ decays to $\overline D^{(*)}D^{(*)}K$ are reconstructed 
exclusively and the corresponding branching fractions or limits are 
determined. The branching fractions of the $B^0$ and of the $B^+$ to 
$\overline D^{(*)}D^{(*)}K$ are found to be   
\begin{center}
$\BR(B^0 \rightarrow \overline D^{(*)} D^{(*)} K) = \left (4.3 \pm 0.3 \stat \pm 0.6 \syst 
\right )\%$, \\
$\BR(B^+ \rightarrow \overline D^{(*)} D^{(*)} K) = \left (3.5 \pm 0.3 \stat \pm 0.5 \syst 
\right )\%$.
\end{center}
A search for decays to orbitally excited $D_s$ states, 
$B \rightarrow \overline D^{(*)} D_{sJ}^{+}$ ($D_{sJ}^{+} \rightarrow D^{(*)0}K^+$) 
is also performed.

\end{abstract}

\pacs{13.25.Hw, 12.15.Hh, 11.30.Er}

\maketitle

\newpage

\setcounter{footnote}{0}
\section{Introduction}
\label{sec:Introduction}
The inconsistency between the measured $\b \to \c \cbar \s$ rate and the
rate of semileptonic $B$ decays has been a long-standing problem in $B$
physics.
Until 1994, it was believed that the $\b \to \c \cbar \s$ transition was
dominated by decays $B \to D_s X$, with some smaller contributions
from decays to charmonium states and to charmed strange baryons.  Therefore,
the $\b \to \c \cbar \s$ branching fraction was computed from the inclusive
$B \to D_s \, X $, $B \to (\c\cbar)\, X $, and $B \to \Xi_c \, X $ branching
fractions, leading to
$\BR(\b \to \c \cbar \s)=(15.8 \pm 2.8)\%$ \cite{browder2}.
Theoretical calculations are unable to simultaneously describe this low
branching fraction and the semileptonic branching fraction of the $B$
meson \cite{bigi}.

As a possible explanation of this problem, it has been
conjectured \cite{buchalla} that $\BR(\b \to \c \cbar \s)$ is
larger and that decays of the type $B \to \Dbar^{(*)} D^{(*)} K\,(X)$
(where $D^{(*)}$ can be either a \Dz, \Dstarz, \Dp, or $D^{*+}$) 
could contribute significantly to the decay rate.
This might also include possible decays to orbitally-excited $D_s$ mesons,
$B \to \Dbar^{(*)} D_{sJ}$, followed by $D_{sJ}\to D^{(*)} K$.
Experimental evidence in support of this picture has been published in the
past few years. This evidence includes the measured branching fraction for
wrong-sign $D$ production, averaged over charged and neutral $B$ mesons,
by CLEO \cite{cleoupv} [$\BR(B \to D\, X)=(7.9\pm 2.2)\% $], and
the observation of a small number of fully reconstructed  decays
\btoddk, both by CLEO \cite{cleoddk} and ALEPH \cite{alephddk}. More recently,
\babar\  \cite{babarddk} and Belle \cite{belleddk} have reported some
preliminary  results on the evidence for transitions
$\Bz \to \Dstarm D^{(*)0} \Kp$ with much larger data sets.

\begin{figure}[tb]
\begin{center}
\epsfig{file=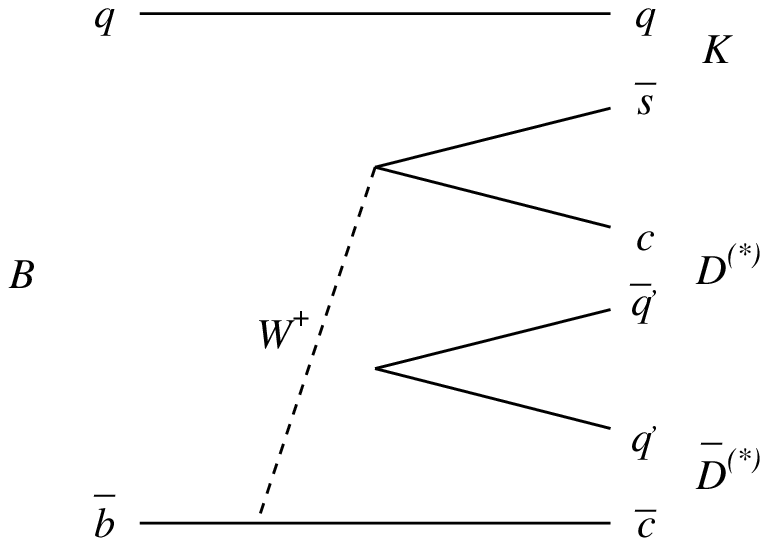,width=7.cm}
\epsfig{file=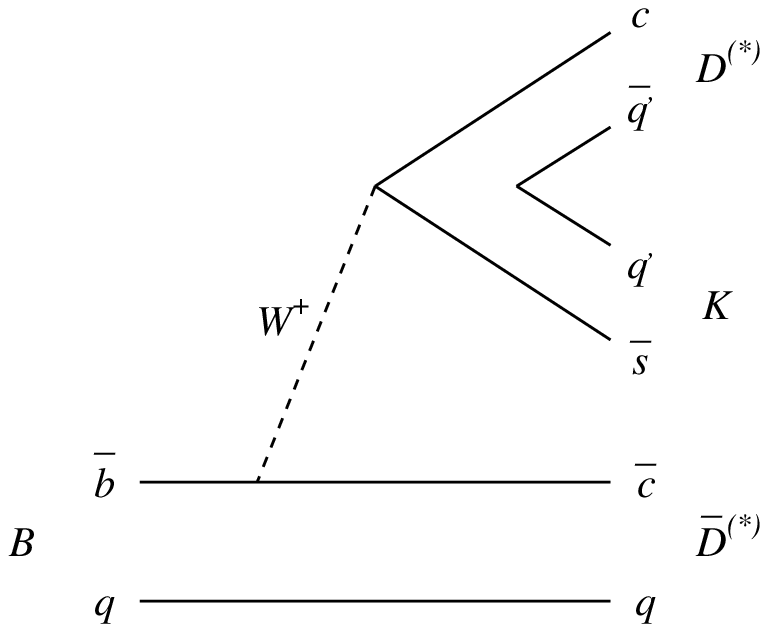,width=7.cm}
\caption{Top: internal \W-emission diagram for the decays 
 $B \to \Dbar^{(*)} D^{(*)} K$. 
Bottom: external \W-emission diagram for the decays
$B \to \Dbar^{(*)} D^{(*)} K$. }
\label{Fi:diagrams}
\end{center}
\end{figure}

\btoddk\ decays can proceed through two different amplitudes: external
\W-emission amplitudes and internal \W-emission amplitudes (also called
color-suppressed amplitudes). Some decay modes proceed purely through one 
of these amplitudes while others can proceed through 
both. Figure~\ref{Fi:diagrams} shows
the possible types for \btoddk\ decays. In \babar, the large
data sets now available allow comprehensive investigations of these
transitions. In this paper, we present measurements
of or limits on the branching fractions for all the possible
$B\to \Dbar^{(*)} D^{(*)} K^0$ and
$B\to \Dbar^{(*)} D^{(*)} \Kp$ decay modes, using events in
which both $D$ mesons are fully reconstructed. Charge conjugate reactions  
are assumed throughout this paper and branching fractions are averaged 
accordingly.
\section{\boldmath The \babar\ detector and dataset}
\label{sec:babar}
The study reported here uses 75.9\invfb\ of data collected at the
\FourS resonance  
with the \babar\ detector at the \pep2\ asymmetric-energy $B$ factory, 
corresponding to $(82.3 \pm 0.9)\times 10^6$ 
\BB pairs.

The \babar\ detector is a large-acceptance solenoidal
spectrometer (1.5$\mathrm{\,T\kern -0.1em}$) 
described in detail elsewhere~\cite{ref:babar}.
The analysis described below makes use of charged track and
\piz reconstruction and charged particle identification.
Charged particle trajectories are measured by a 5-layer double-sided
silicon vertex tracker (SVT) and a 40-layer drift chamber (DCH), which also
provide ionization measurements (\dedx) used for particle
identification. For charged tracks with $p>1 \gevc$, the measured transverse momentum 
with respect to the beam axis ($p_T$) has a resolution $\sigma_{p_T}$ 
such that 
\begin{equation}
\frac{\sigma_{p_T}}{p_T} = 0.13 \% p_{T} + 0.45 \%,
\end{equation}
where $p_T$ is measured in ${\mathrm{Ge\kern -0.1em V\!/}c}$.

Photons and electrons are measured in the electromagnetic
calorimeter (EMC). The EMC consists of 6580 Thallium-doped CsI crystals, 
which instrument the barrel and forward endcap; the crystals are
arranged
in a quasi-projective geometry. The electromagnetic
calorimeter resolution $\sigma_E$ can be expressed as 
\begin{equation}
\frac{\sigma_E}{E} = \frac{2.3\%}{E^{\frac{1}{4}}}
\oplus 1.9 \%,
\end{equation}
where the energy $E$ is measured in \gev.

Charged particle identification is provided by the average energy loss (\dedx) 
in the tracking devices and by an internally reflecting ring-imaging Cherenkov
detector (DIRC). The DIRC comprises 144 quartz bars, divided into 12 
sectors, which transport the Cherenkov light to a water-filled
expansion volume equipped with 
10751 photomultiplier tubes. A $K/\pi$ separation better than four standard 
deviations is achieved for momenta below 3\gevc.

\section{\boldmath ${B}$ candidate selection}
\label{sec:Analysis}
The \Bz and \Bu mesons are reconstructed in a sample of hadronic events
for all the possible $\Dbar D K$ modes, namely
$\Bz\to D^{(*)-}D^{(*)0}\Kp$,
$D^{(*)-}D^{(*)+}\Kz$, $\Dbar^{(*)0}D^{(*)0}\Kz$ and
$\Bu\to \Dbar^{(*)0}D^{(*)+}\Kz$,
$\Dbar^{(*)0}D^{(*)0}\Kp$, $D^{(*)-}D^{(*)+}\Kp$.
$\Kz$ mesons are reconstructed only from the decays $\KS\to\pipi$.
To eliminate the background from
continuum $\epem\to\qqbar$ events, we require that the ratio of
the second to zeroth Fox-Wolfram moments of the event \cite{ref:fox} 
be less than 0.45.

The \KS candidates are reconstructed from two oppositely charged tracks
consistent with coming from a common vertex and having an invariant mass
within $\pm 9\mevcc$ of the nominal \KS mass. For most of the channels
involving a \KS, we require that the \KS vertex be displaced from the
interaction point for the event by at least 0.2\cm in the plane
transverse to the beam axis direction.
  The \piz candidates are reconstructed from pairs of photons, each with
 energy greater than 30\mev, which are required to have an invariant mass
$115<m_{\gaga}<150\mevcc$. The \piz from a \Dstarz decay must
have momentum between 70\mevc and 450\mevc in the \FourS frame,
while the \piz from $\Dz\to \Km\pip\piz$ must have energy 
greater than 200\mev in the laboratory frame.

The $D^*$ candidates are reconstructed in the decay modes
$D^{*+}\to \Dz\pip$, $D^{*+}\to D^+\piz$,  $\Dstarz\to \Dz\piz$, and
$\Dstarz\to \Dz\g$. A $\pm 3 \sigma$ interval around the nominal mass 
difference $\Delta m= m(\Dstar)-m(\Dz)$ is used to select \Dstar mesons, 
where $\sigma$ is the measured mass difference resolution and is equal to
$1 \mevcc$ for $\Dstarp \to \Dz\pip$ and $\Dstarp\to \Dp \piz$ decays, 
$1.3 \mevcc$ for $\Dstarz\to\Dz\piz$ decays, and $3.3 \mevcc$ for 
$\Dstarz\to\Dz\g$ decays. The mode $D^{*+}\to D^+\piz$ is used only 
in the reconstruction of decays $\Bz \to D^{*-}D^{*+}\KS$ and 
$\Bu \to D^{*-}D^{*+}\Kp$.

The \Dz and \Dp mesons are reconstructed in the decay modes
$\Dz\to \Km\pip$, $\Km\pip\piz$, $\Km\pip\pim\pip$, and
$\Dp\to \Km\pip\pip$, by selecting track
combinations with invariant mass within $\pm 2\sigma$ of the average
measured $D$ mass.  The average $D$ mass and the $D$ mass resolution
$\sigma$ used in this selection are fitted from the data itself,
using a large inclusive sample of $D$ decays. The resolution is
equal to $7\mevcc$ for \dkpi\ decays, $13 \mevcc$ for \dkpipiz\ decays,
$5.7 \mevcc$ for \dkpipipi\ decays, and $5.5 \mevcc$ for $\Dp\to\Km\pip\pip$
decays.
For modes involving two \Dz mesons, at least one of them is required to
decay to $\Km \pip$, except for the decay modes $D^{*-}D^{*+}\KS$,
$D^{*-}D^{*+}\Kp$, and $D^{*-}\Dz\Kp$, which have lower background and
for which all combinations are accepted.
All $K$ and $\pi$ tracks are required to be well
reconstructed in the tracking detectors and to originate from a common vertex.
Charged kaon identification, based on the measured Cherenkov angle in the DIRC
and the \dedx\ measurements in the drift chamber and the vertex tracker, is
used for most $D$ decay modes, as well as for the $\Kp$ from the
$B$ meson decay.

$B$ candidates are reconstructed by combining one $\Dbar^{(*)}$, one $D^{(*)}$
and one $K$ candidate. A mass-constrained kinematic fit is applied to all 
intermediate
particles (\Dstarz, $D^{*+}$, \Dz, \Dp, \KS, \piz).
Since the $B$ mesons are produced via \epem
$\to$ \upsbb, the energy of the $B$ meson in the \FourS rest frame
is given by the beam energy in the center-of-mass frame, $\sqrt{s}/2$,
which is known much more
precisely than the energy of the $B$ candidate. Therefore, to isolate the
$B$ meson signal, we use two kinematic variables: the difference
between the reconstructed energy of the $B$ candidate and the beam energy in
the center of mass frame (\de), and the beam energy substituted mass (\mes),
defined as
\begin{equation}
\mes = \sqrt{\left(\frac{\sqrt{s}}{2}\right)^{2}-p_B^{*2}},
\end{equation}
where $p_B^*$ is the momentum of the reconstructed $B$ in
the \FourS frame. Signal events have \mes close to the nominal $B$ meson mass,
 5.279\gevcc, and \de\ close to  0\mev. Due to imperfect modeling of the 
charged $K$ energy loss in the detector material, the central value of \de\
is slightly shifted away from 0\mev by an amount $\Delta E_{\mathrm{shift}}=(-5 \pm 1)\mev$, 
which is fitted from the data themselves 
(Figs.~\ref{fig:figsum} (a),~\ref{fig:figsum} (b)). 
When several candidates are selected in the same event,
only the candidate with the lowest $|\Delta E-\Delta E_{\rm shift}|$  
value is considered (``best candidate'').
From Monte Carlo studies, this algorithm is found to give the best
reconstruction efficiency and the lowest cross-feed rate between the different
$\Dbar^{(*)}D^{(*)}K$ modes; it is found to introduce no bias
on the signal extraction, since the latter is performed from the \mes\
spectra only. However, in Fig.~\ref{fig:figsum}, to avoid the bias on
$\Delta E$ inherent to this method, \de\ spectra are shown without applying 
this selection.

\section{Evidence for \boldmath $B \to \Dbar^{(*)}D^{(*)}K$}
\label{sec:Sumresults}
\begin{figure*}
\epsfig{file=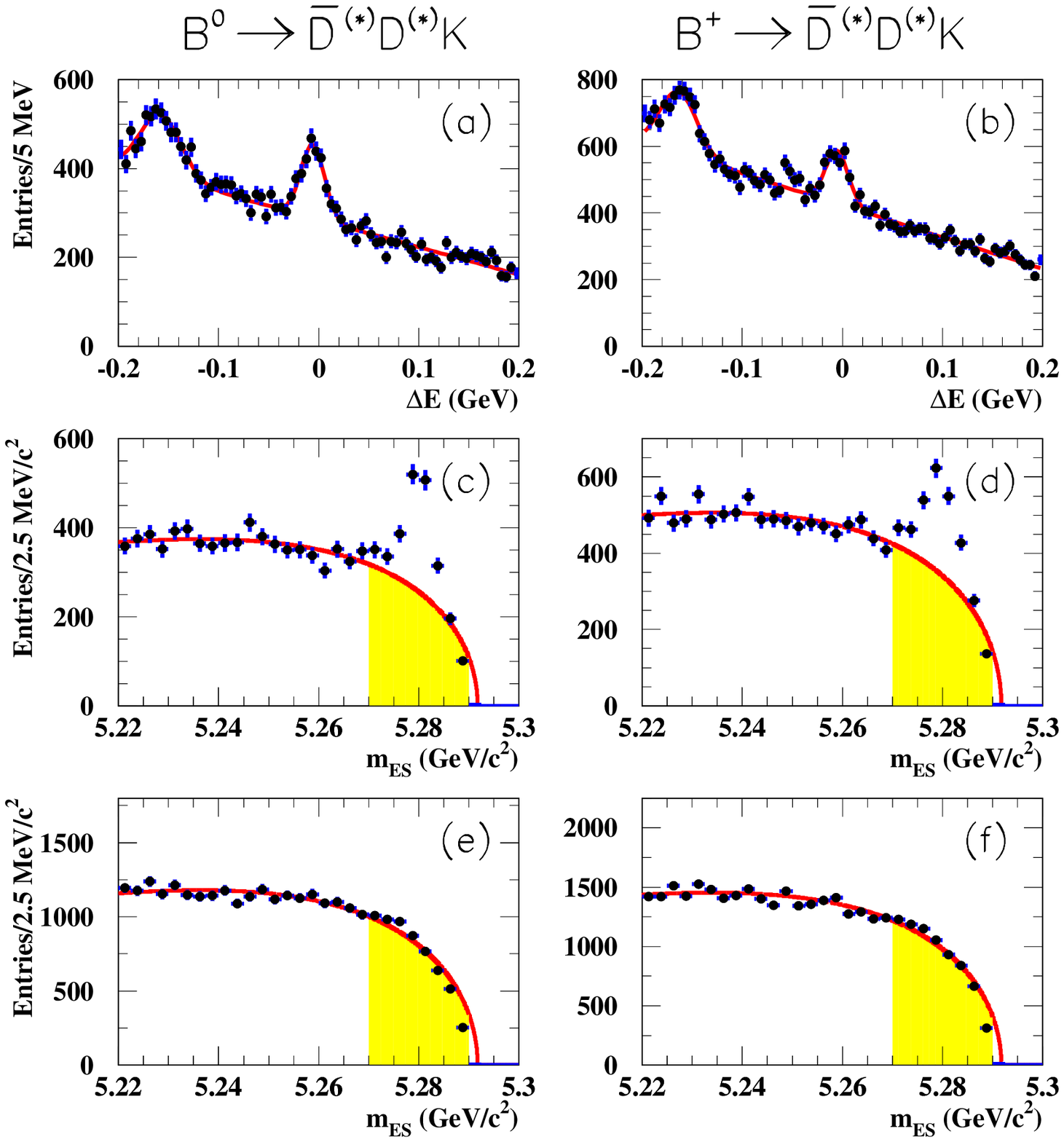,width=0.8\textwidth}
\begin{center}
\caption{The $\Delta E$ and $m_{\rm ES}$ spectra (a,c,e) for the sum of all
the $\Bz\to\Dbar^{(*)}D^{(*)}K$ modes and (b,d,f) for the sum of all the
$\Bu\to\Dbar^{(*)}D^{(*)}K$  modes.
 (a,b): $\Delta E$ for $5.27<m_{\rm ES}<5.29\gevcc$.
 (c,d): $m_{\rm ES}$ for 
$\left| \Delta E  - \Delta E_{\mathrm{shift}} \right| <2.5\sigma_{\de}$.
 (e,f): $m_{\rm ES}$  for $\Delta E>50\mev$ (background control region).
 The curves superimposed on the $m_{\rm ES}$ spectra correspond to the
 background fits described in the text and the shaded regions
represent the background in the signal region $5.27<m_{\rm ES}<5.29\gevcc$.}
\label{fig:figsum}
\end{center}
\end{figure*}

The \mes\ and \de\ spectra of the selected events are shown in
Fig.~\ref{fig:figsum} for the sum of all the decay modes, separately for
\Bz and \Bu. The \de\ spectra are shown for
events in the signal region defined by $5.27<m_{\rm ES}<5.29\gevcc$.
Signal events appear in the peak near 0\mev\ when reconstructed correctly,
while the peak around $-160$\mev\ is due to
$\Dbar^* D K$  and $\Dbar D^* K$ decays reconstructed as
$\Dbar D K$ and to $\Dbar^* \Dstar K$ decays reconstructed as 
$\Dbar^* D K$ or $\Dbar D^* K$.
The \mes\ spectra for the signal region are shown for events with \de\ within
$\pm 2.5\sigma_{\Delta E}$ of the central $\Delta E$ value for the signal. The
resolution $\sigma_{\Delta E}$ is determined from the data and is equal to
9.9\mev\ for events involving no \Dstarz and 11.3\mev\ for events involving
one \Dstarz. For events with two \Dstarz candidates, the resolution is 
estimated from 
the Monte Carlo simulation to be 13.8\mev. As explained above, 
only the candidate with the lowest $|\Delta E - \Delta E_{\mathrm{shift}}|$ 
appears in the \mes\ spectra in case of multiple candidates. 
Both the \mes\ spectra for the \de\ signal region and
the \de\ spectra show clear evidence of a signal.
On the contrary, the \mes\ spectra for the background control region
$\Delta E>50\mev$ do not contain any excess of events in the $B$ signal 
region as expected. 
When fitting the \mes\ spectra, the combinatorial background component is 
empirically described by a threshold function \cite{ref:argus} (henceforth 
referred to as the ARGUS distribution),  
\begin{eqnarray}
{ \frac{{\rm d}N}{{\rm d}m_{\rm ES}} } & = 
& f( m_{\rm ES};A,\zeta) = 
A  m_{\rm ES} \sqrt{1-\frac{m_{\rm ES}^2}
{m_0^{2}}} \\
\nonumber        &  & \times  
\exp{\left[-\zeta \left( 1-\frac{m_{\rm ES}^2}
{m_0^{2}}\right)\right]},
\end{eqnarray}
where $m_0$ represents the kinematic upper limit and is held fixed at the 
center-of-mass beam energy $E^*_{\mathrm beam} = 5.291 \gev$, and $A$ is a 
normalization factor. The function depends on a free parameter
$\zeta$ that is determined from a fit to the \mes\ spectrum
of the background control region. The number of combinatorial background
events in the signal region is then estimated by normalizing the ARGUS
distribution to the region $5.22<m_{\rm ES}<5.27\gevcc$ in the $\Delta E$ slice 
containing the signal (Figs.~\ref{fig:figsum}(c), ~\ref{fig:figsum}(d)) and
extrapolating it to the signal region $5.27<m_{\rm ES}<5.29\gevcc$.
The fitted ARGUS distributions are overlaid on the \mes\ spectra of
Fig.~\ref{fig:figsum}. 

The number of background events predicted in the signal region by the fit
 is 1889$\pm$24 for neutral $B$ mesons and 2512$\pm$27 for charged $B$ mesons,
 while 2712 and 3482 events are observed, giving an excess of 823$\pm$57 \Bz 
and 970$\pm$65 \Bu events in the signal region.

\begin{figure*}
\begin{tabular}{lr}
\hspace{-1cm}
\epsfig{file=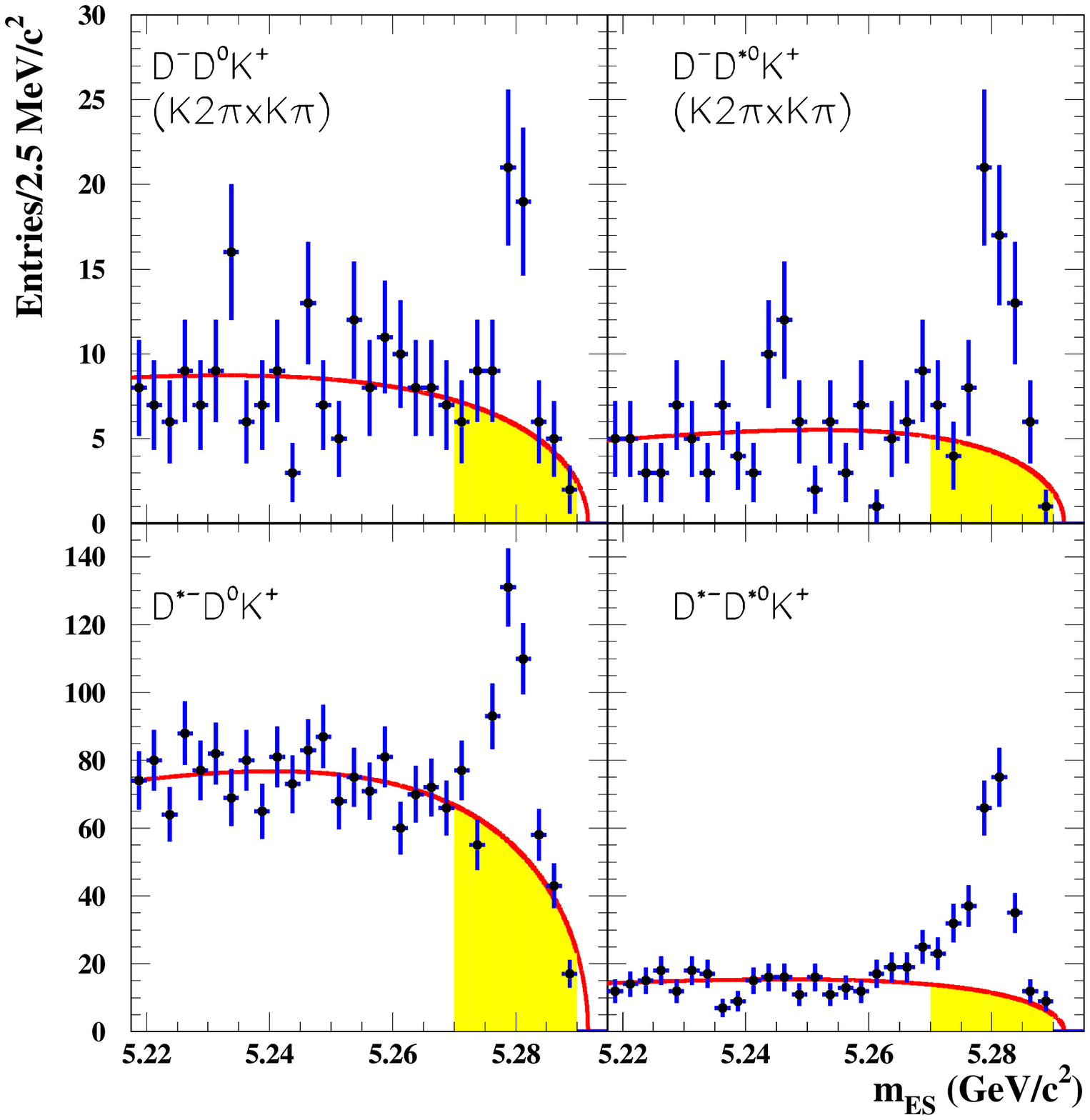,width=0.55\textwidth} &
\hspace{-1.5cm}
\epsfig{file=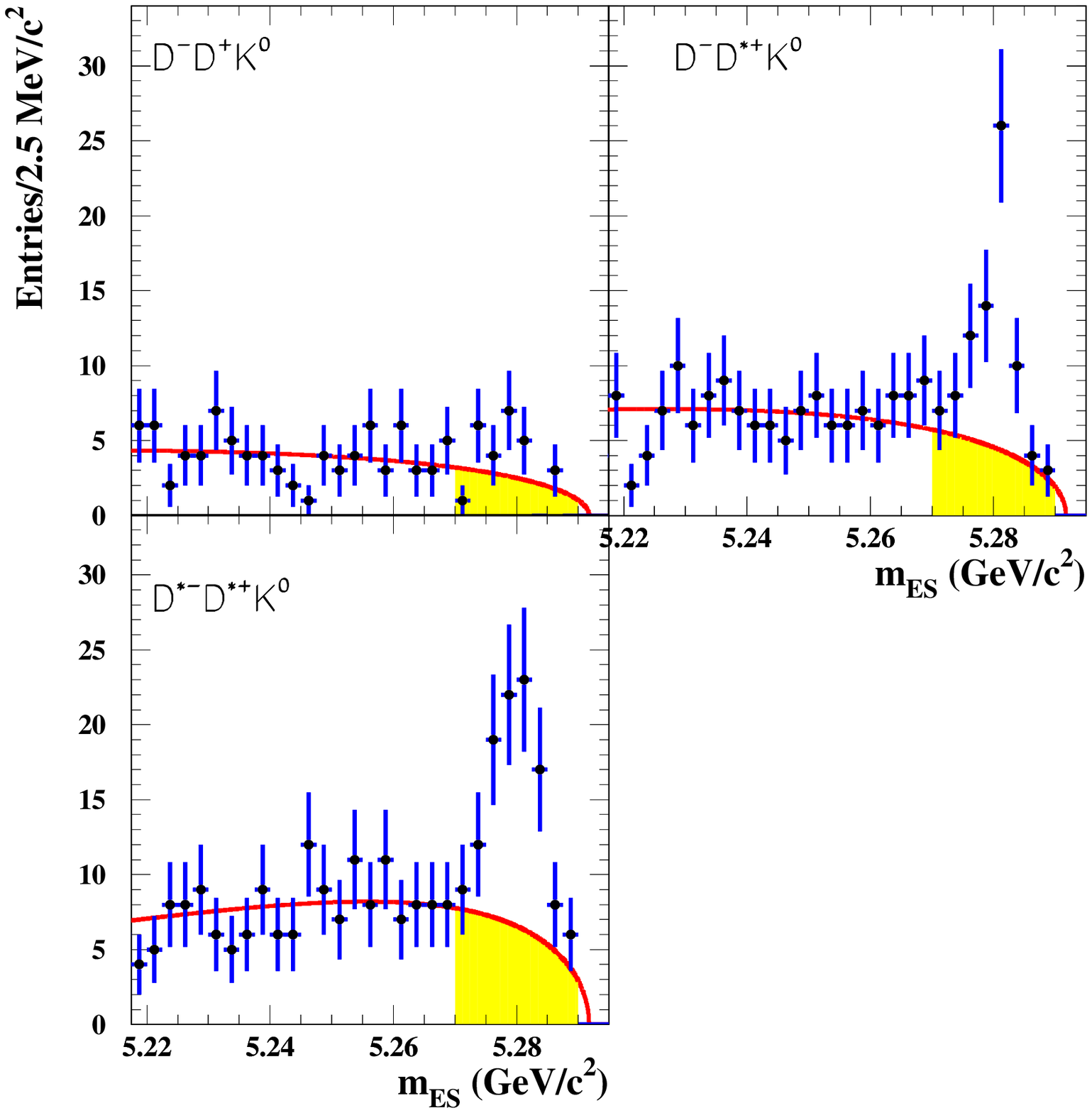,width=0.55\textwidth} \vspace{-1cm} \\
\hspace{-1cm}
\epsfig{file=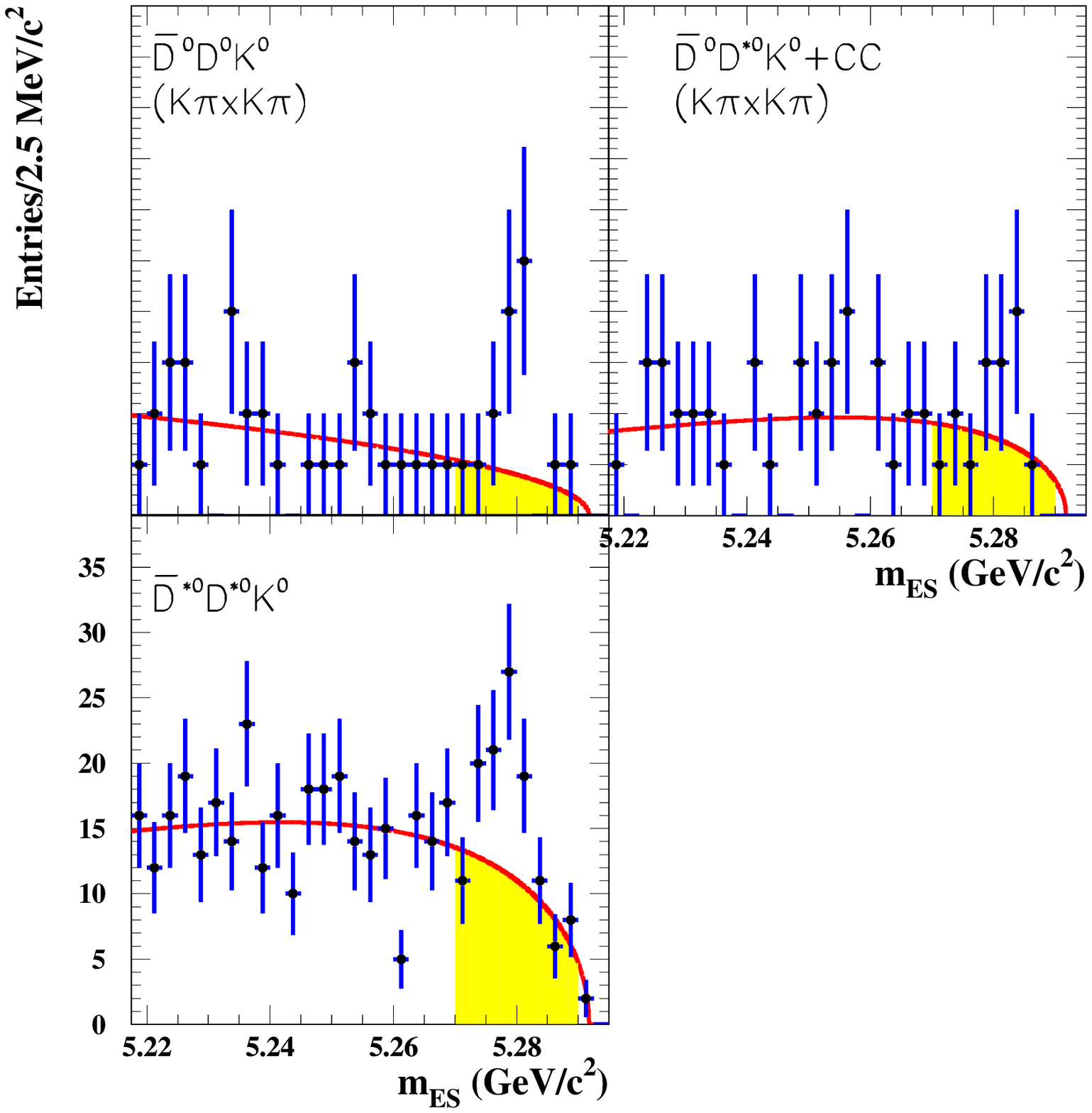,width=0.55\textwidth} & \\
\end{tabular}
\vspace{-1cm}
\begin{center}
\caption{The $m_{\rm ES}$ spectra of the ten
$\Bz\to\Dbar^{(*)} D^{(*)}K$ modes. For each mode, all the $D$
decay submodes used in the analysis have been summed, except for
$B$ modes for which the $\Dbar \times D$ decay mode is listed
explicitly on the plot.
The curves correspond to the background fits described in the text and the
shaded regions represent the background in the signal region.
Upper left: pure external \W-emission (spectator)
decays $\Bz\to D^{(*)-}D^{(*)0} \Kp$.
Upper right: external+internal \W-emission decays 
$\Bz\to D^{(*)-} D^{(*)+}\KS$.
Lower left: pure internal \W-emission (color-suppressed) decays $\Bz\to
\Dbar^{(*)0} D^{(*)0} \KS$.}
\label{fig:b0modes}
\end{center}
\end{figure*}
\begin{figure*}
\begin{tabular}{lr}
\hspace{-1cm}
\epsfig{file=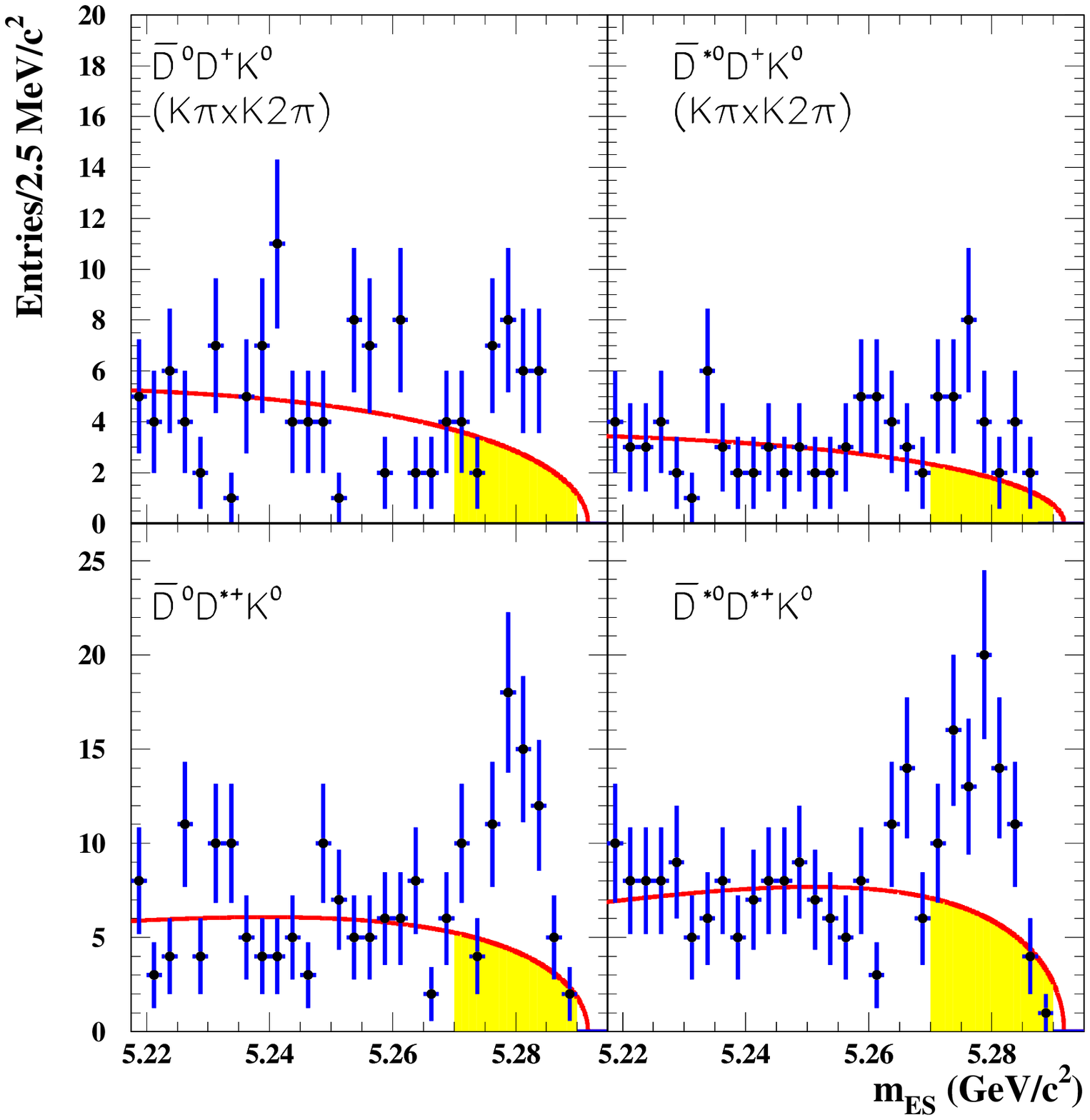,width=0.55\textwidth} &
\hspace{-1.5cm}
\epsfig{file=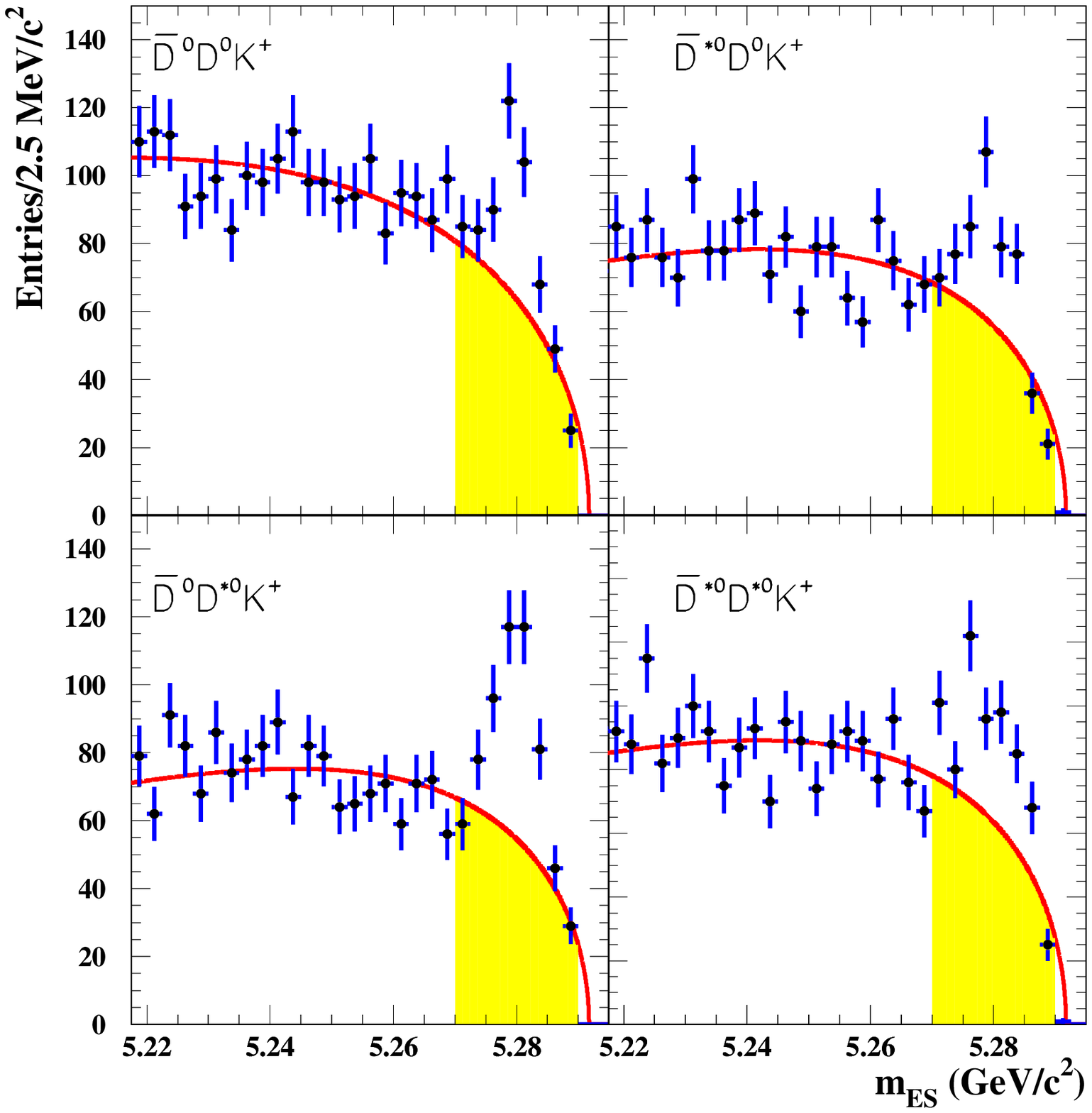,width=0.55\textwidth} \vspace{-1cm} \\
\hspace{-1cm}
\epsfig{file=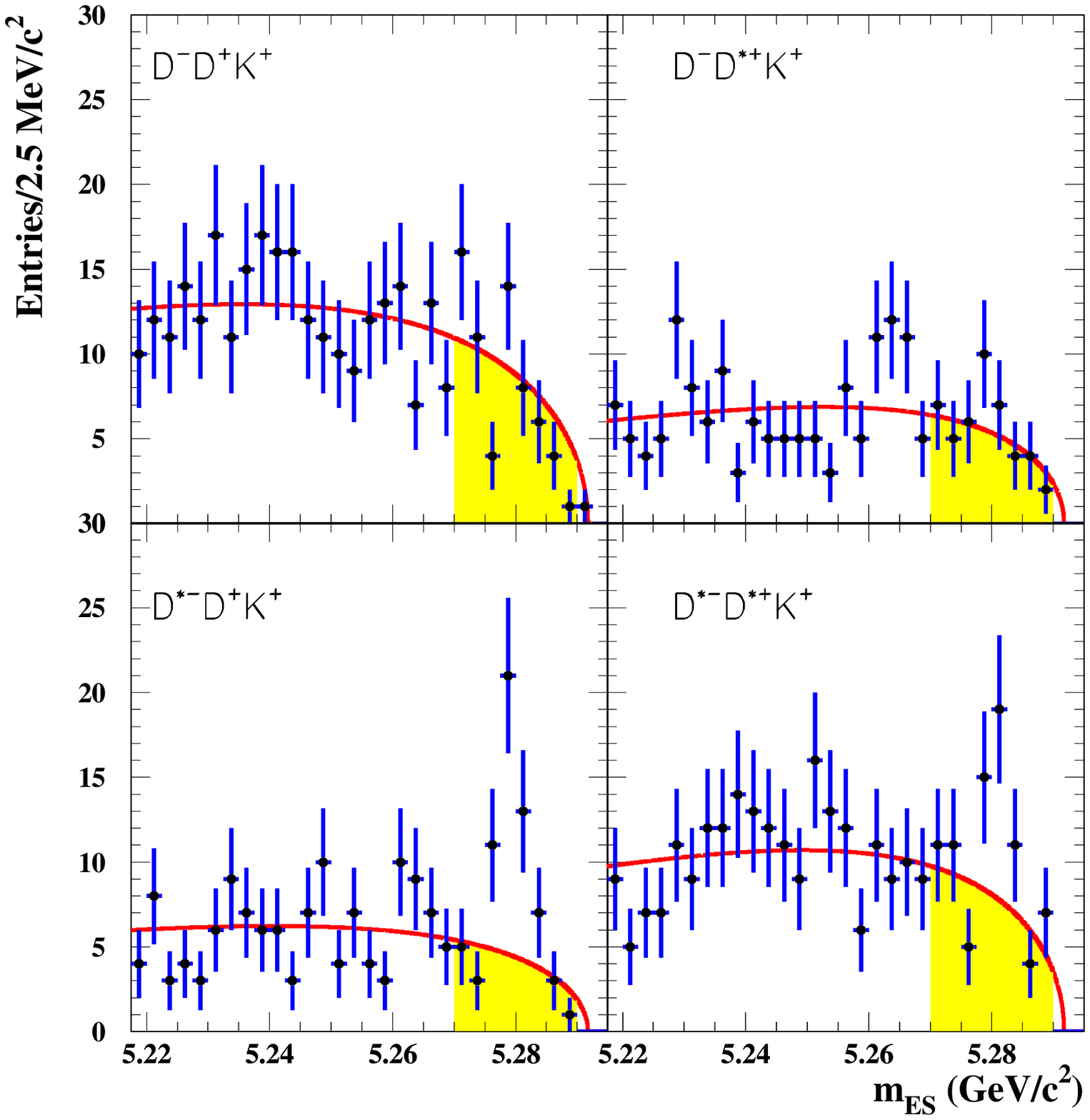,width=0.55\textwidth} & \\
\end{tabular}
\vspace{-1cm}
\begin{center}
\caption{The $m_{\rm ES}$ spectra of the twelve $\Bu\to\Dbar^{(*)}\Dstar K$
modes.
For each mode, all the $D$ decay submodes used in the analysis have been
summed, except for $B$ modes for which the $\Dbar \times D$ decay mode
is listed explicitly on the plot.
 The curves correspond to the background fits described in
the text and the shaded regions represent the background in the signal region.
Upper left: pure external \W-emission (spectator)
 decays $\Bu\to\Dbar^{(*)0} D^{(*)+}\KS$.
Upper right: external+internal \W-emission decays 
$\Bu\to\Dbar^{(*)0} D^{(*)0}\Kp$.
Lower left: pure internal \W-emission (color-suppressed) decays 
$\Bu\to D^{(*)-} D^{(*)+}\Kp$.}
\label{fig:bchmodes}
\end{center}
\end{figure*}

\section{Determination of branching fractions}
\label{sec:exclusive}
\begin{table*}[tb]
\caption{Number of events and branching fractions for each mode. 
The excess (column 4) is the 
difference between the total yield observed in the signal region 
$5.27<\mes<5.29\gevcc$ and the 
combinatorial background. It includes the contribution from the signal 
itself and from the 
cross-feed from the other $\Dbar^{(*)}D^{(*)}K$ modes. The number of 
cross-feed events, 
computed from the cross-feed matrix and from the measured 
$B \rightarrow \Dbar^{(*)}D^{(*)}K$ branching fractions, 
is given in column 5. When omitted, the predicted number of cross-feed 
events is smaller than 5 and has been 
neglected in the branching fraction calculations, as explained in the text.   
The first error on each branching fraction is the statistical uncertainty 
and the second
one is the systematic uncertainty. For the decay modes with a significance 
$S/\sqrt{B}$ 
 smaller than 4, a 90\% confidence level (C.L.) upper limit is also 
derived. Here, $B$ is the sum of the combinatorial background and of
cross-feed, while 
$S=N-B$, where $N$ is the total yield in the signal region. The decay mode 
$\Bu\to\Dstarzb\Dz\Kp$ has a large number of signal events but its 
significance is lower than 4 because of the large cross-feed from 
$\Bu\to \Dzb \Dstarz\Kp$ and $\Bz\to D^{*-} \Dz\Kp$. The fractional statistical error on 
the branching fractions cannot be directly related to the fractional statistical error on
 the excess since the different decay submodes of the $\bar D D$ pair (not detailed in the table) 
enter with different statistical weights in the branching fraction calculation, while the yields 
given here are a raw sum over all the $\bar D D$ decay submodes.  
The statistical uncertainty on the background, dominated by the uncertainty on
 the ARGUS shape parameter $\zeta$, is incorporated in the systematic error on
 the branching fractions. }
\begin{center}
\begin{tabular}{|l|c|c|c|c|c|c|} \hline
              & Total yield & Estimated      & Excess =   &  & Branching & 90\%~C.L.\\
  $B$ decay   &  $N$ in the & combinatorial  & Signal +   & cross-& fraction  & upper   \\
    mode    & signal region & background     & cross-feed & feed  & $(\%)$& limit $(\%)$  \\ \hline
\multicolumn{7}{|c|}{\Bz decays through external \W-emission amplitudes} \\ \hline
 $\Bz\to \Dm \Dz\Kp$      & 599 & 479$\pm$12 & 120$\pm$27 & - & $0.17 \pm 0.03 \pm 0.03$ &\\
$\Bz\to\Dm\Dstarz\Kp$     & 468 & 337$\pm$10 & 131$\pm$24 & - & $0.46 \pm 0.07 \pm 0.07$ &\\
$\Bz\to D^{*-} \Dz\Kp$    & 584 & 399$\pm$11 & 185$\pm$27 & - & $0.31^{+0.04}_{-0.03}\pm 0.04$ &\\
$\Bz\to D^{*-}\Dstarz\Kp$ & 289 & 84$\pm$5   & 205$\pm$18 & - & $1.18 \pm 0.10 \pm 0.17 $&\\ \hline
\multicolumn{7}{|c|}{\Bz decays through external+internal \W-emission amplitudes} \\ \hline
$\Bz\to\Dm\Dp\Kz$         &  26 & 19$\pm$2   &   7$\pm$5  & - & $0.08^{+0.06}_{-0.05}\pm 0.03$ & $0.17$\\
$\Bz\to D^{*-}\Dp\Kz+\Dm D^{*+}\Kz$ & 84 & 34$\pm$3 &50$\pm$10 & - & $0.65 \pm 0.12 \pm 0.10$ & \\
$\Bz\to D^{*-} D^{*+}\Kz$ & 116 & 48$\pm$4   &  68$\pm$11 & - & $0.88^{+0.15}_{-0.14}\pm 0.13$ &\\ \hline
\multicolumn{7}{|c|}{\Bz decays through internal \W-emission amplitudes} \\ \hline
$\Bz\to \Dzb \Dz \Kz$     & 175 & 173$\pm$7  & 2$\pm$15   & - & $0.08 \pm 0.04 \pm 0.02$ & $0.14$ \\
$\Bz\to\Dzb \Dstarz \Kz+ \Dstarzb \Dz \Kz$ & 248 & 225$\pm$8 & 23$\pm$18 & - & $0.17^{+0.14}_{-0.13}\pm 0.07$ & $0.37$ \\
$\Bz\to \Dstarzb \Dstarz \Kz$& 123 &  81$\pm$6 & 42$\pm$13 & 19.8 & $0.33^{+0.21}_{-0.20}\pm 0.14$ & $0.66$ \\ \hline
\multicolumn{7}{|c|}{\Bu decays through external \W-emission amplitudes} \\ \hline
$\Bu\to \Dzb \Dp\Kz$ & 367 & 317$\pm$9      & 50$\pm$21 & -   & $0.18 \pm 0.07 \pm 0.04$ & $0.28$ \\
$\Bu\to \Dstarzb\Dp\Kz$ & 216 & 175$\pm$7   & 41$\pm$16 & 9.6 & $0.41^{+0.15}_{-0.14}\pm 0.08$ & $0.61$ \\
$\Bu\to \Dzb D^{*+}\Kz$ &  77 & 31$\pm$3    & 46$\pm$9  & -   & $0.52^{+0.10}_{-0.09}\pm 0.07$ &\\
$\Bu\to \Dstarzb D^{*+}\Kz$ & 89 & 43$\pm$4 & 46$\pm$10 & 9.0 & $0.78^{+0.23}_{-0.21}\pm 0.14$ & \\ \hline
\multicolumn{7}{|c|}{\Bu decays through external+internal \W-emission amplitudes} \\ \hline
$\Bu\to \Dzb \Dz \Kp$     & 627 & 469$\pm$11 &158$\pm$27   & -    & $0.19 \pm 0.03 \pm 0.03$       &\\
$\Bu\to\Dstarzb\Dz\Kp$ & 552 & 411$\pm$11 &141$\pm$26      & 75.3 & $0.18^{+0.07}_{-0.06} \pm 0.04$ & $0.38$ \\
$\Bu\to \Dzb \Dstarz\Kp$ & 623 & 402$\pm$11 &221$\pm$27    & 37.1 & $0.47 \pm 0.07 \pm 0.07$       &\\
$\Bu\to\Dstarzb\Dstarz\Kp$ & 675 & 468$\pm$15 & 207$\pm$30 & 66.6 & $0.53^{+0.11}_{-0.10} \pm 0.12$ &\\  \hline
\multicolumn{7}{|c|}{\Bu decays through internal \W-emission amplitudes} \\ \hline
$\Bu\to \Dm\Dp\Kp$ & 64 & 65$\pm$4 & -1$\pm$9         &-& $0.00 \pm 0.03 \pm 0.01$ & $0.04$ \\
$\Bu\to\Dm D^{*+}\Kp$ & 45 & 39$\pm$4 & 6$\pm$8       &-& $0.02 \pm 0.02 \pm 0.01$ & $0.07$ \\
$\Bu\to D^{*-} \Dp\Kp$ & 64 & 32$\pm$3 & 32$\pm$9     &-& $0.15 \pm 0.03 \pm 0.02$ &       \\
$\Bu\to D^{*-} D^{*+}\Kp$ & 83 & 60$\pm$4 & 23$\pm$10 &-& $0.09 \pm 0.04 \pm 0.02$ & $0.18$ \\ \hline
\end{tabular}
\end{center}
\label{tab:ddkyields}
\end{table*}

In the following, the subscript $k$ will be used to identify the different
\btoddk\ decay modes (i.e., $\Dzb \Dz\Kp$, $D^{*-}\Dz\Kp$, ...).
The subscript $i$ will be used to identify the different decay submodes of the
$\Dbar D$ pair (i.e., $i$ =$ K\pi\times K\pi$, $K\pi\times K\pi\piz$,
$K\pi\times K3\pi$, ...). The subscript $ik$ will therefore refer to
$B$ mode $k$ decaying into $\Dbar D$ submode $i$.

The \mes\ spectra obtained after a $\pm 2.5 \sigma_{\Delta E}$ selection on
$\left(\Delta E-\Delta E_{\mathrm{shift}}\right)$ for all the different $\Dbar^{(*)} D^{(*)} K$ modes are shown in
Fig.~\ref{fig:b0modes} (\Bz decay modes) and Fig.~\ref{fig:bchmodes}
(\Bu decay modes).
The corresponding event yields, computed as explained below, are given in
Table~\ref{tab:ddkyields}. In Figs.~\ref{fig:b0modes} and \ref{fig:bchmodes}
and in Table~\ref{tab:ddkyields}, for a given $B$ decay mode the signals
from the different $\Dbar D$ decay submodes have been summed.
However, to take advantage of the different signal-to-background ratios
of the various submodes, the information from each submode is entered
separately in a likelihood function used to calculate the
\btoddk\ branching fractions.
As a first step, the ARGUS distribution shape parameter of each submode,
$\zeta_{ik}$, is determined from a maximum likelihood fit to the \mes\ spectra
of the background control region $\Delta E>50\mev$. An ARGUS distribution with
the shape parameter $\zeta$ fixed to this value $\zeta_{ik}$ is then fitted
to the \mes\ distribution for the signal region
$|\Delta E-\Delta E_{\mathrm{shift}}|<2.5\sigma_{\Delta E}$,
excluding from the fit events with $5.27 < m_{\rm ES} < 5.29\gevcc$.
The factor $A_{ik}$ is calculated so that the function is normalized
to the total number of background events and
the number of background events, $\mu^{\rm bkg}_{ik}$, in the signal region
for this submode is calculated as
\begin{equation}\label{Eq:nbkg}
\mu^{\rm bkg}_{ik} = \int_{5.27}^{5.29} f\left(x;A_{ik},\zeta_{ik}\right) 
{\rm d}x.
\end{equation}
If $n_k$ submodes are used for a given mode, the branching fraction for
that mode is then extracted by maximizing the following likelihood:
\begin{equation}\label{Eq:poisson}
L_k = \prod_{i=1}^{n_k} \frac{{\mu_{ik}}^{N_{ik}} e^{-\mu_{ik}}}{N_{ik}!},
\end{equation}
where $N_{ik}$ and $\mu_{ik}$ are the observed and predicted number of events,
respectively, in the signal region. 
$\mu_{ik}$ is the sum of three contributions:
\begin{itemize}
\item the predicted signal $\mu_{ik}^S$, which is the product of
the (unknown) branching fraction $\BR_k$ of decay mode $k$,
the reconstruction efficiency $\epsilon_{ik}$, the intermediate branching
fractions
$\BR_{i}^{\Dbar D}$, and the number of \BB events, $N_{\BB}$, assuming
that the number of \BzBzb meson pairs produced at the \FourS resonance 
is equal to the number of \BpBm pairs:
\begin{equation}
\mu_{ik}^S = \BR_k \times N_{\BB} \times \epsilon_{ik} 
\times \BR_{i}^{\Dbar D};
\end{equation}

\item the number of combinatorial background events, $\mu^{\rm bkg}_{ik}$,
determined as described above (Eq.~\ref{Eq:nbkg});
\item the peaking background $\mu^{\rm peak}_{ik}$ from other \btoddk\ decay
modes. The cross-feed between different $\Dbar D$ decay submodes is found to 
be negligible and $\mu^{\rm peak}_{ik}$ is therefore calculated as

\begin{equation}
\mu_{ik}^{\rm peak} = \sum_{l \ne k} \BR_l \times N_{\BB} \times
\epsilon '(il\to ik)  \times \BR_{i}^{\Dbar D},
\end{equation}
where $\epsilon '(il\to ik)$ is the cross-feed matrix element that represents
the probability for $B$ mode $l$ to be reconstructed as $B$ mode $k$
for $\Dbar D$ decay submode $i$. The only 
significant cross-feed is observed between decay modes where a fake 
\Dstarz replaces a real \Dstarp or a real \Dz, for instance between 
$D^{*-}\Dz\Kp$ and $\Dstarzb \Dz \Kp$, or between $\Dstarzb \Dz \Kp$ and
 $\Dzb \Dstarz \Kp$. 
\end{itemize}

The branching fractions $\BR_k$ for the sets of decay modes that have 
significant cross-feed are simultaneously fit, by maximizing the product 
$\prod_{k}L_k$ of the corresponding likelihood functions. 

The \Dstar and $D$ branching fractions used in the branching fraction
calculation are summarized in Table \ref{tab:brpdg} \cite{ref:pdg}.
Branching fractions for decay modes reconstructed with a \KS are calculated
for neutral $K$ mesons, including \KL.
The selection efficiencies and the cross-feed matrices for each mode are
obtained from a detailed Monte Carlo simulation, in which the detector
response is modeled with the Geant4 program \cite{ref:geant4}.
The simulated event samples of \btoddk\ decays used for the efficiency
calculation are generated according to a phase space model. 
For each decay submode, samples of about 15000 signal events
have been produced. In addition, data are used whenever
possible to determine detector performance: tracking efficiencies are
determined by identifying tracks in the silicon vertex detector and measuring
the fraction that is well reconstructed in the drift chamber; the kaon
identification efficiency is estimated from a sample of
$\Dstarp\to\Dz\pip$, $\Dz\to \Km\pip$ decays; the $\g$
and $\piz$ efficiencies are measured by comparing the ratio of events,
$N(\taup \to \nutb h^+\piz)/N(\taup \to \nutb h^+\piz\piz)$,
to the published branching fractions \cite{cleotau}.
Typical efficiencies range from 20\%
for \bdzdzk\ with both \Dz mesons decaying to $\Km\pip$,  to less than 1\%
for \btodsdsk\ ($D^{*+} \to \Dz \pip$, $D^{*-} \to \Dzb \pim$ )
with \Dz mesons decaying to $\Km\pip\piz$ or $\Km\pip\pim\pip$.

\begin{table}[tb]
\caption{Submode branching fractions used in the analysis \cite{ref:pdg}.
The errors on $\BR(\Dz\to \Km \pip \piz)$ and $\BR(\Dz \to \Km \pip \pim
\pip)$
correlated to the error on $\BR(\Dz \to \Km \pip)$ are indicated
separately with the subscript $K\pi$.}
\begin{center}
\begin{tabular}{|l|c|}
\hline
\textbf{Mode} & $\BR$ (\%)  \\
\hline\hline
$\Dz\to K^- \pi^+$ & $3.80\pm 0.09\ $ \\
$\Dz\to K^- \pi^+ \pi^0$ &$13.10 \pm 0.84 \pm 0.31_{K\pi}$ \\
$\Dz \to K^- \pi^+ \pi^- \pi^+$ & $7.46  \pm 0.30 \pm 0.18_{K\pi}$\\ \hline
$\Dp\to K^- \pi^+ \pi^+$ & $9.1 \pm 0.6\ $ \\ \hline
$D^{*+}\to \Dz \pi^+$ & $67.7 \pm 0.5\ $ \\
$D^{*+}\to \Dp \piz$ & $30.7 \pm 0.5\ $ \\ \hline
$\Dstarz\to \Dz \piz$ & $61.9 \pm 2.9\ $ \\
$\Dstarz\to \Dz \gamma$ &$38.1 \pm 2.9\ $ \\ \hline
$\KS\to \pipi$&$68.60 \pm 0.27\ $ \\ \hline
\end{tabular}
\end{center}
\label{tab:brpdg}
\end{table}

\section{Systematic studies}
\label{sec:Systematics}
\begin{table*}[htb]\caption{Fractional systematic uncertainties
on efficiencies and branching fractions.}
\begin{center}
\begin{tabular}{|l|l|}
\hline
Item & Fractional uncertainty on efficiency or branching fraction  \\
\hline\hline
Charged track reconstruction & 0.8\% per track for tracks with more that 12
hits required in the Drift
Chamber \\
                       & 1.2\% per track for tracks without Drift Chamber
requirement \\ \hline
\KS reconstruction        & 2.5\% per \KS, added in quadrature to the
track reconstruction error \\ \hline
\piz reconstruction & 5.1\% per \piz   \\ \hline
$\g$ from $\Dstarz \to \Dz \g$ & 5.1\% per $\g$ (correlated with the
\piz systematic) \\ \hline
\Kpm identification & 2.5\% per \Kpm  \\ \hline
Vertex         & 1.3\%  per 2 track vertex  \\
reconstruction & 3.1\%  per 3 track vertex  \\
               & 5.7\%  per 4 track vertex  \\ \hline
$\sigma(\Delta E)$ & 2\% for modes with zero or one \Dstarz \\
                   & 5\% for modes with two \Dstarz's \\ \hline
Background description & 5\% to 20\% (ARGUS shape parameter $\zeta$, mode dependent) \\ 
                       & 3.5\%  (end point $m_0$)      \\ \hline 
Monte Carlo statistics & 2\% to 10\% per $\Dbar D$ submode
(mode and submode dependent) \\ \hline
Intermediate br. fraction        & see Table \ref{tab:brpdg}  \\ \hline
Number of \BB   & 1.1\%  \\ \hline
Decay model & 5\%  \\ \hline
\end{tabular}
\end{center}
\label{tab:systematics}
\end{table*}
Due to the large number of \Kpm mesons and to the large track multiplicities involved in
the decays \btoddk, the dominant systematic uncertainties come from our level
of understanding of the charged kaon identification and of the charged-particle 
tracking efficiencies. Both systematic uncertainties are estimated for each track
and are given in Table \ref{tab:systematics}. Another important systematic is 
 the uncertainty linked to the background description. One of its components is from the 
 uncertainty on the number of background events and is dominated by the 
uncertainty on the ARGUS shape parameter $\zeta$. The relative error on the 
branching fractions associated with this component varies from 5\% up to 20\%
depending on the mode and is uncorrelated from one mode to another.  
The other component is from the end point $m_0$ of the ARGUS distribution. Changing $m_0$ by $\pm 1\mev$ 
 results in a $\pm 1.4\%$ change of the fitted combinatorial background. The 
associated fractional error on each branching fractions is estimated to be 
$\pm 3.5\%$ in average and is correlated between all the modes.  
Other systematic uncertainties are due to uncertainties on the $D$ and \Dstar branching
fractions, the \piz reconstruction efficiencies, the $D$ vertex fit quality
requirements, and the $\Delta E$ resolution used to define the
signal box, as well as  the statistical uncertainty on the efficiency due to the finite
size of the Monte Carlo simulation samples and the uncertainty on
the number of \BB events in the data sample. The different contributions 
to the systematic uncertainties on  the branching fractions
are summarized in Table \ref{tab:systematics}.

Possible decay model dependences of the efficiencies were also studied by
generating the decays $\Bz \to D^{*-}D_{s1}^+$ and
$\Bz \to D^{*-}D'^+_{s1}$
($D^+_{s1}, D'^+_{s1}\to \Dstarz\Kp)$,
where $D^+_{s1}$ is the narrow ($\Gamma=1\mev$, $m=2535.35\mevcc$) orbitally
excited $1^+$ state of the $D_{sJ}$ system and $D'^+_{s1}$ is a wide
($\Gamma=250\mev$, $m=2560\mevcc$)
$D_{sJ}$ resonance. The efficiency for reconstructing these
modes was compared to the efficiency found for
$\Bz \to D^{*-} \Dstarz\Kp$ decays generated with a phase-space model.
We found no statistically significant difference in efficiencies; we assign
a systematic uncertainty equal to the statistical error 
of the ratio (5\%).

\section{search for resonant substructure}

$B\to \Dbar^{(*)} D^{(*)0} \Kp$ decay modes are used to probe the
possible presence of intermediate $D_{sJ}$ resonances decaying
into $D^{(*)0}\Kp$, where $D_{sJ}$ are P-wave 
excitations of the $c \bar s$ system. In the heavy-quark (charm) mass limit, 
the spin of the heavy quark decouples, and both the spin $J$ of the meson and 
the total angular momentum (spin plus orbital) $j_q$ of the light quark become 
good quantum numbers  \cite{ref:rosner, ref:isgur}. There are four P-wave 
states with the following spin-parity and 
light-quark angular momenta: $0^+$ ($j_q=1/2$), $1^+$ ($j_q=1/2$), 
$1^+$ ($j_q=3/2$), $2^+$ ($j_q=3/2$). The two $j_q=3/2$ states can only 
undergo D-wave decay, and therefore have narrow widths. The remaining $j_q=1/2$
 states decay via S-waves and are expected to be quite broad. Their masses are
 predicted to be $\approx 2.48 \gevcc$ ($0^+$) and $\approx 2.55 \gevcc$ 
($1^+$), while their widths are predicted to be a few hundred $\mev$ 
\cite{ref:godfrey}. However, the recent observation by the \babar\  
collaboration of a narrow state decaying to $D_{s}^+\piz$, with a mass of 
$(2316.8 \pm 0.4) \mevcc$ (statistical error only) \cite{ref:dsj2517}, would 
contradict these predictions  and could indicate that the $J^P=0^+$ state has 
a mass lower than the $D^{(*)}K$ 
threshold; if this interpretation is confirmed, the $0^+$ state would 
therefore not contribute to the \btoddk\ final state.
   
In the analysis described below, the two narrow resonances, $D_{s1}^+(2536)$ 
and $D_{sJ}^+(2573)$, are considered. The full Dalitz plot for the decay $\Bz 
\to D^{*-} \Dstarz\Kp$ is also examined. 

\subsection{\boldmath $D_{s1}^+(2536)$}
$D_{s1}^+(2536)$ is the most probable resonance to contribute to
\btoddk\ decays. It has already been observed and its measured
parameters are $m=(2535.35\pm 0.60)\mevcc$, $\Gamma<2.3 \mev$, $J^P=1^+$,
and $j_q=\frac{3}{2}$ \cite{ref:pdg}. Because of conservation of 
parity and angular momentum, only the decays $D_{s1}^+(2536)\to D^*K$
are allowed. In this analysis, a search is made for the $D_{s1}^+(2536)$  in  
the final state $\Dstarz\Kp$ in the four decay
modes $\Bz\to\Dm\Dstarz\Kp$, $\Bz\to\Dstarm\Dstarz\Kp$, $\Bu\to
\Dzb\Dstarz\Kp$, and $\Bu\to\Dstarzb\Dstarz\Kp$. This resonance is not
reconstructed in the $\Dstarp \KS$ final state due to its lower reconstruction 
efficiency. 

\begin{figure}[htb]
\begin{center}
\epsfig{file=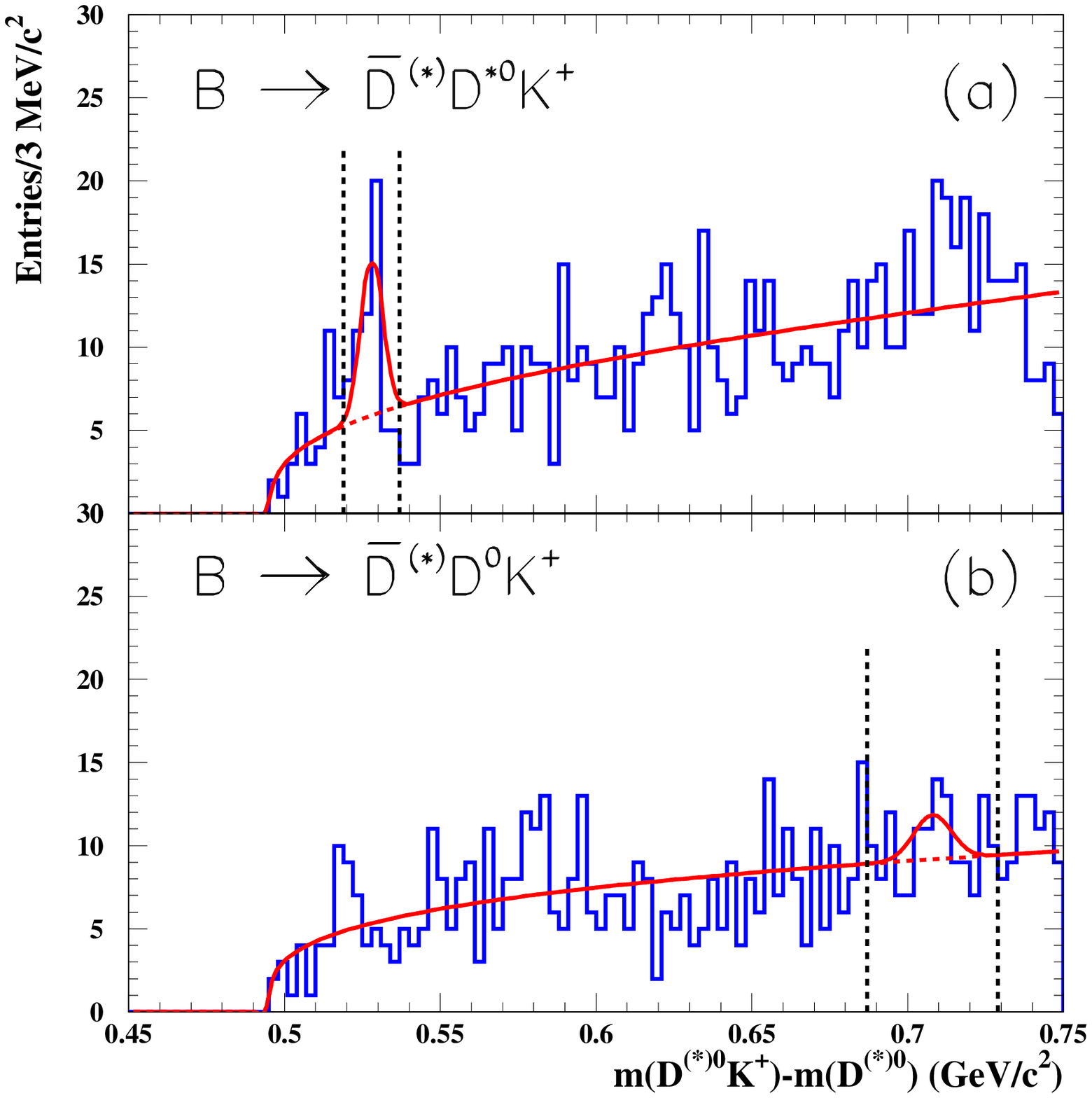,width=9cm} \caption{(a) $\Delta
m = m\left(\Dstarz\Kp\right)-m\left(\Dstarz\right)$ distribution
for events reconstructed in the $B\to\Dbar^{(*)}\Dstarz\Kp$ signal
regions.  (b) $\Delta m = m\left(\Dz\Kp\right)-m\left(\Dz\right)$
distribution for events reconstructed in the
$B\to\Dbar^{(*)}\Dz\Kp$ signal regions.} \label{fig:d2s-fig1}
\end{center}
\end{figure}

Figure~\ref{fig:d2s-fig1}(a) shows the distribution of the variable
$\Delta m = m\left(\Dstarz\Kp\right)-m\left(\Dstarz\right)$ for
the events reconstructed in the signal region 
($5.27 < \mes < 5.29 \gevcc$) for
these four decay modes. The distribution is fitted with a Gaussian
function describing the signal. The combinatorial background is
represented by a threshold function defined as 
\begin{equation}\label{Eq:threshold}
g\left(\Delta m\right) = N\left(\Delta m-\Delta m_0\right)^\beta
e^{\alpha\left(\Delta m-\Delta m_0\right)}.
\end{equation}
The parameters of the Gaussian function (mean value $\Delta m_1=527.9 \mevcc$ and 
standard deviation $\sigma_{\Delta m}=3.5 \mevcc$) are fixed to the values obtained 
from a fit to the same distribution resulting from the reconstruction of 
inclusive $D_{s1}^+(2536)\to\Dstarz\Kp$ decays in a large sample of 
events. This procedure yields an estimated signal of
$28_{-7}^{+8}$ $D_{s1}^+(2536) \to \Dstarz\Kp$ events out 
of $764 \pm 50$ $B\to \Dbar^{(*)} \Dstarz \Kp$ events. 

In order to extract upper limits on the contribution of $D_{s1}^+(2536)$ to 
$\btoddk$ decays, 
the same method is applied to the four individual decay modes, as 
shown in Fig.~\ref{fig:d2s-fig2}(a). The region $519 < \Delta m < 537 \mevcc$, 
illustrated by the dashed lines in Fig.~\ref{fig:d2s-fig1}(a), is defined as 
the signal region and the number of combinatorial background events in this
region is estimated from the fit by integrating the background function $g$
defined in Eq.~\ref{Eq:threshold}. The total number of events observed 
in the signal box is compared to the expected combinatorial background when 
extracting the limits. Table~\ref{tab:ds1} summarizes the results obtained 
and gives a 90 \% confidence level (C.L.) upper limit on the quantity $R_{2536}$, 
\begin{small}
\begin{equation}
R_{2536} = \frac{n(B\to\Dbar^{(*)}D_{s1}^+(2536)) \times
\BR\left(D_{s1}^+(2536)\to\Dstarz\Kp\right)}
{n\left(B\to\Dbar^{(*)}\Dstarz\Kp\right)}.
\end{equation}
\end{small}
where \begin{small} $n(B\to\Dbar^{(*)}D_{s1}^+(2536)) \times
\BR\left(D_{s1}^+(2536)\to\Dstarz\Kp\right) $ \end{small} and 
\begin{small} $n(B\to\Dbar^{(*)}\Dstarz\Kp)$ 
\end{small} are the observed number of signal events.
Using the $B\to\Dbar^{(*)}\Dstarz\Kp$ branching fraction measurements of 
Table~\ref{tab:ddkyields}, these results can be compared to the only existing
 measurement of inclusive $D_{s1}^+(2536)$ production in $B$ decays, 
$\BR\left( B \rightarrow D_{s1}^+(2536) X\right)<0.95\%$ at 90\% 
C.L. \cite{ref:cleo2536}.  

\begin{table}[htb]
\caption{$D_{s1}^+(2536)\to\Dstarz\Kp$ contributions to 
$B\to \Dbar^{(*)} \Dstarz \Kp$ decays.}
\begin{center}
\begin{tabular}{|l|c|c|c|} \hline
                  & Total yield         &            & $R_{2536}$  \\
  $B$ decay mode  & in $D_{s1}^+(2536)$ & Estimated  & 90\% C.L. \\
                  & signal region       & background & upper limit \\ \hline
    $\Dm\Dstarz\Kp$      & 16 &  $7.8\pm 0.6$& $11.6\%$ \\
    $\Dstarm\Dstarz\Kp$  & 13 &  $7.3\pm 0.6$&  $5.8\%$ \\
    $\Dzb\Dstarz\Kp$     & 12 & $11.1\pm 0.8$&  $3.1\%$ \\
    $\Dstarzb\Dstarz\Kp$ & 20 &  $8.7\pm 0.5$&  $9.1 \%$ \\ \hline
\end{tabular}
\end{center}
\label{tab:ds1}
\end{table}

\subsection{\boldmath $D_{sJ}^+(2573)$}
\begin{figure*}
\begin{tabular}{lr}
\hspace{-1cm}
\epsfig{file=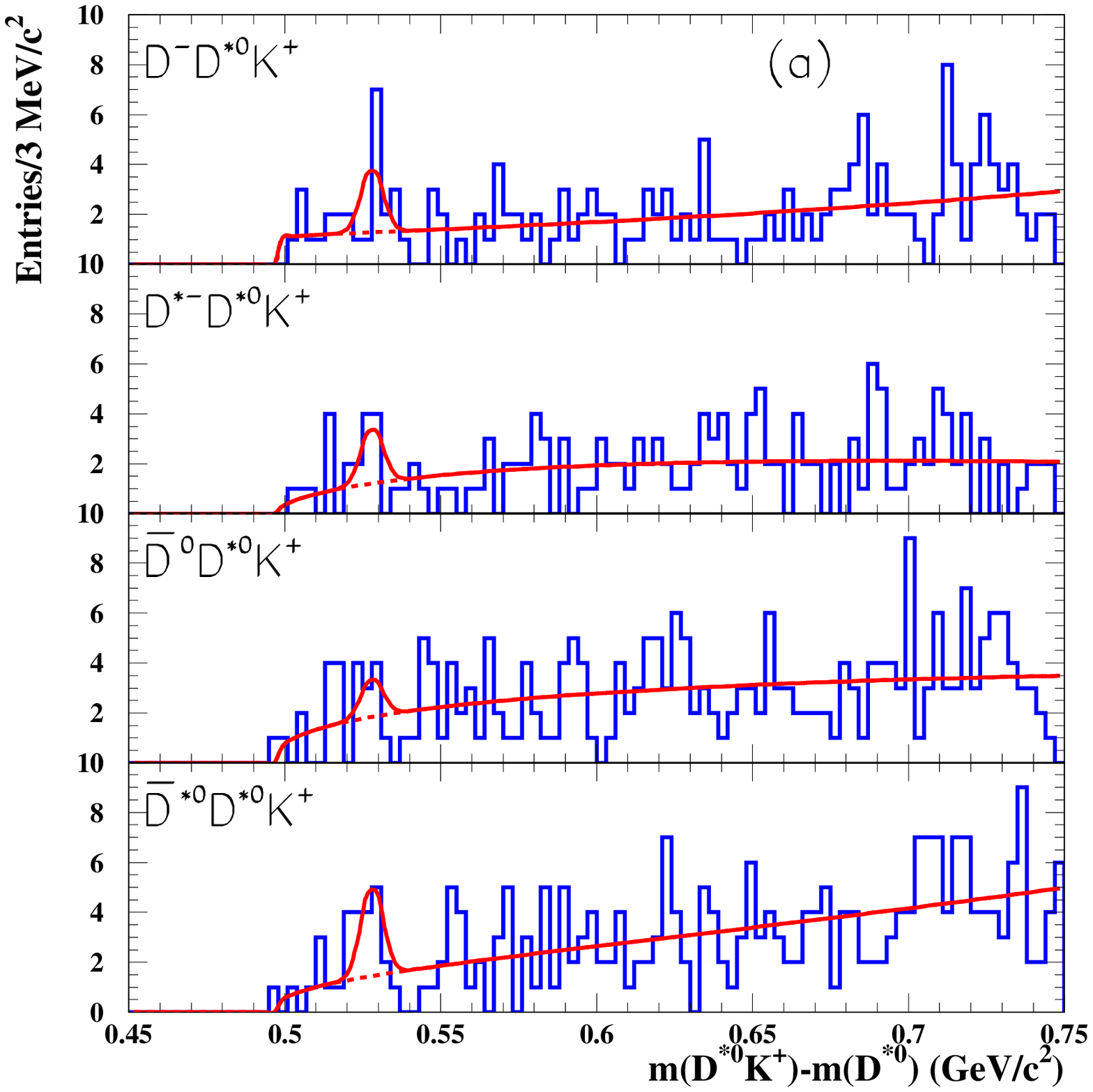,width=0.55\textwidth} &
\hspace{-1.5cm}
\epsfig{file=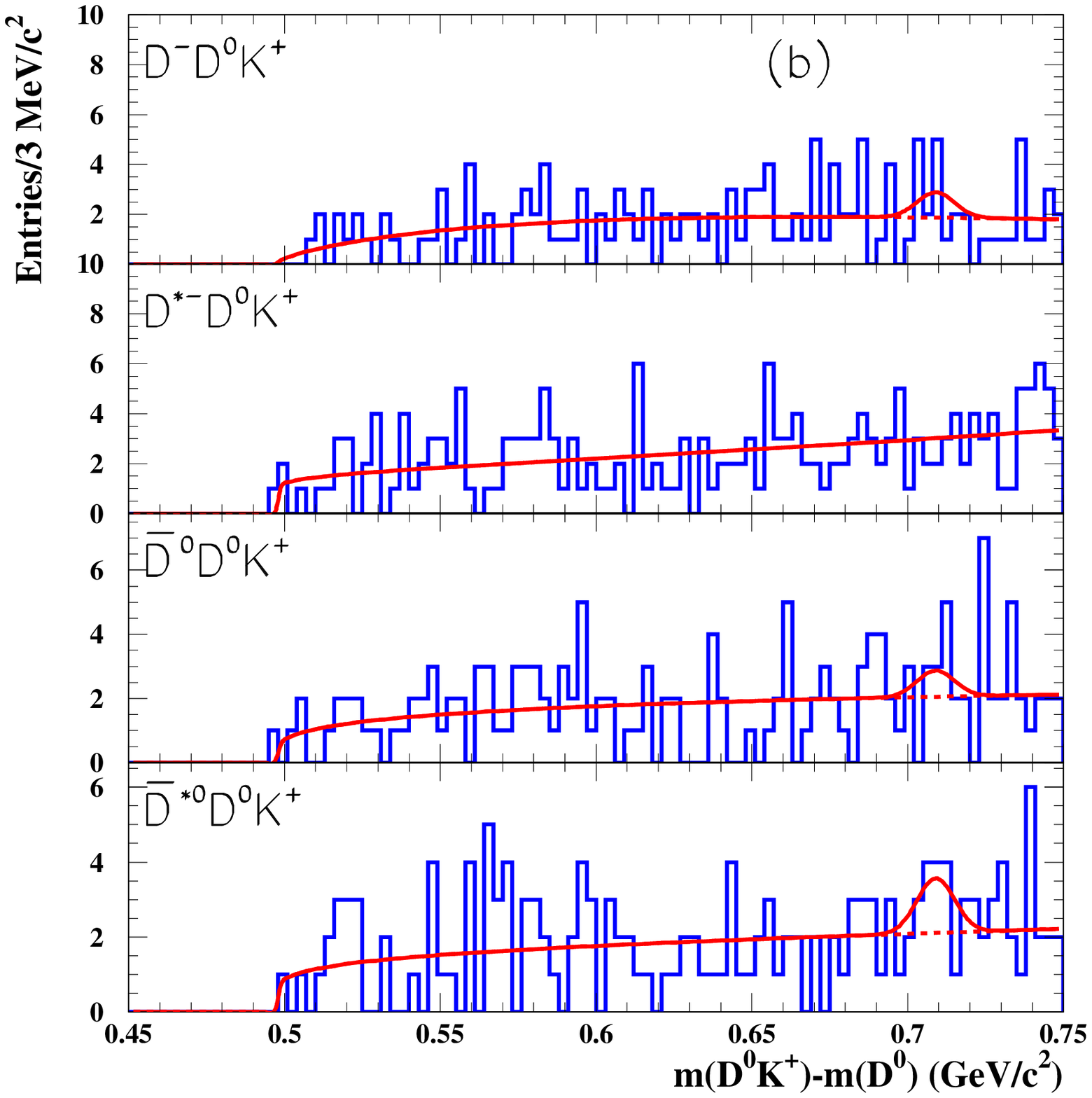,width=0.55\textwidth}  \\
\end{tabular}
\begin{center}
\caption{ (a) $\Delta m = m\left(\Dstarz\Kp\right)-m\left(\Dstarz\right)$ 
distributions of the events reconstructed in the four 
$B\to \Dbar^{(*)}\Dstarz\Kp$ decay modes, with $5.27<\mes<5.29\gevcc$. 
(b) $\Delta m = m\left(\Dz\Kp\right)-m\left(\Dz\right)$ 
distributions of the events reconstructed in the four $B\to \Dbar^{(*)}\Dz\Kp$
 decay modes, with $5.27<\mes<5.29\gevcc$. These distributions are fit with 
the sum of a threshold function $f$  (Eq.~\ref{Eq:threshold}) for the 
background components and a Gaussian function for the $D_{s1}(2536)$
and $D_{sJ}(2573)$ components. 
The mean value and standard deviation of the Gaussian distributions have been 
fixed to the values obtained from a fit to the inclusive $D_{s1}(2536)$ and 
$D_{sJ}(2573)$ samples, as described in the text.}
\label{fig:d2s-fig2}
\end{center}
\end{figure*}

The contribution of the $D_{sJ}^+(2573)$ resonance to \btoddk\ decays
is also studied. This resonance is thought to be the other narrow state in 
the $j_q = 3/2$ orbitally excited $D_{sJ}$ doublet, together with 
the $D_{s1}^+(2536)$. The world average values of its mass and width 
are $m=(2573.5\pm 1.7)\mevcc$ and 
$\Gamma = \left(15_{-4}^{+5}\right)\mev$ \cite{ref:pdg}. 
Its spin-parity has not been measured but its natural width and decay 
properties are consistent with a $J^P=2^+$ state \cite{ref:pdg}. If it is 
indeed a spin-2 resonance, it cannot be obtained with a $W$-mediated tree 
diagram but might still be reached through final state interactions. 

The allowed decay modes of the $D_{sJ}^+(2573)$ are $DK$ and $D^*K$, both 
proceeding through a D-wave. Because of the limited phase space, the latter 
is highly suppressed \cite{ref:godfrey}. In this analysis, a search is made 
for the $D_{sJ}^+(2573)$ in the decay mode $\Dz\Kp$, in the four 
channels $\Bz\to\Dm\Dz\Kp$, $\Dstarm\Dz\Kp$, $\Dzb\Dz\Kp$, 
and $\Dstarzb\Dz\Kp$.

The method developed for the $D_{s1}^+(2536)$ study is applied. 
Figure~\ref{fig:d2s-fig1} (b) shows the $\Delta m = m\left(\Dz\Kp\right)
-m\left(\Dz\right)$ distribution for the events reconstructed in all
four $B\to \Dbar^{(*)}\Dz\Kp$ decay modes. The mean value
and the standard deviation of the Gaussian component of the fit function
are fixed, respectively, to $\Delta m = 708 \mevcc$ and $\sigma_{\Delta m} =
6 \mevcc$, which are the values derived from a large inclusive 
$D_{sJ}^+(2573)\to\Dz\Kp$ data sample. The fitted yield of  
$D_{s1}^+(2573) \to \Dz\Kp$ decays is $13\pm 9$  events out 
of $604 \pm 54$ $B\to \Dbar^{(*)} \Dz \Kp$ events.

Defining the signal region $687 < \Delta m < 729 \mevcc$, $90 \%$
C.L. upper limits on the contribution of $D_{sJ}^+(2573)$
to \btoddk\ decays are set for each of the four 
individual decay modes (Fig.~\ref{fig:d2s-fig2}(b)). The number of events
observed in the signal box, the number of background events expected from the 
fits, and the resulting limits are given in Table~\ref{tab:ds2}. 
$R_{2573}$ is defined here as 
\begin{small}
\begin{equation}
R_{2573} = \frac{n(B\to\Dbar^{(*)}D_{sJ}^+(2573)) \times
\BR\left(D_{sJ}^+(2573)\to\Dz\Kp\right)}
{n\left(B\to\Dbar^{(*)}\Dz\Kp\right)}.
\end{equation}
\end{small}

\begin{table}[ht]
\caption{$D_{sJ}^+(2573)\to\Dz\Kp$ contributions to
$B\to \Dbar^{(*)} \Dz \Kp$ decays.}
\begin{center}
\begin{tabular}{|l|c|c|c|} \hline
                     & Total yield         &           & $R_{2573}$  \\
  $B$ decay mode     & in $D_{sJ}^+(2573)$ & Estimated & 90\% C.L. \\
                     & signal region       & background& upper limit \\ \hline
    $\Dm\Dz\Kp$      & 25                  & $26\pm 3$ &   $6.3 \%$ \\
    $\Dstarm\Dz\Kp$  & 41                  & $42\pm 3$ &   $5.0 \%$ \\
    $\Dzb\Dz\Kp$     & 38                  & $29\pm 3$ &  $12.2 \%$ \\
    $\Dstarzb\Dz\Kp$ & 37                  & $30\pm 3$ &  $12.3 \%$ \\ \hline
\end{tabular}
\end{center}
\label{tab:ds2}
\end{table}


\subsection{Dalitz-plot analysis of the decay \boldmath 
${\Bz \to \Dstarm \Dstarz\Kp}$}
\begin{figure*}[tb]
\begin{center}
\epsfig{figure=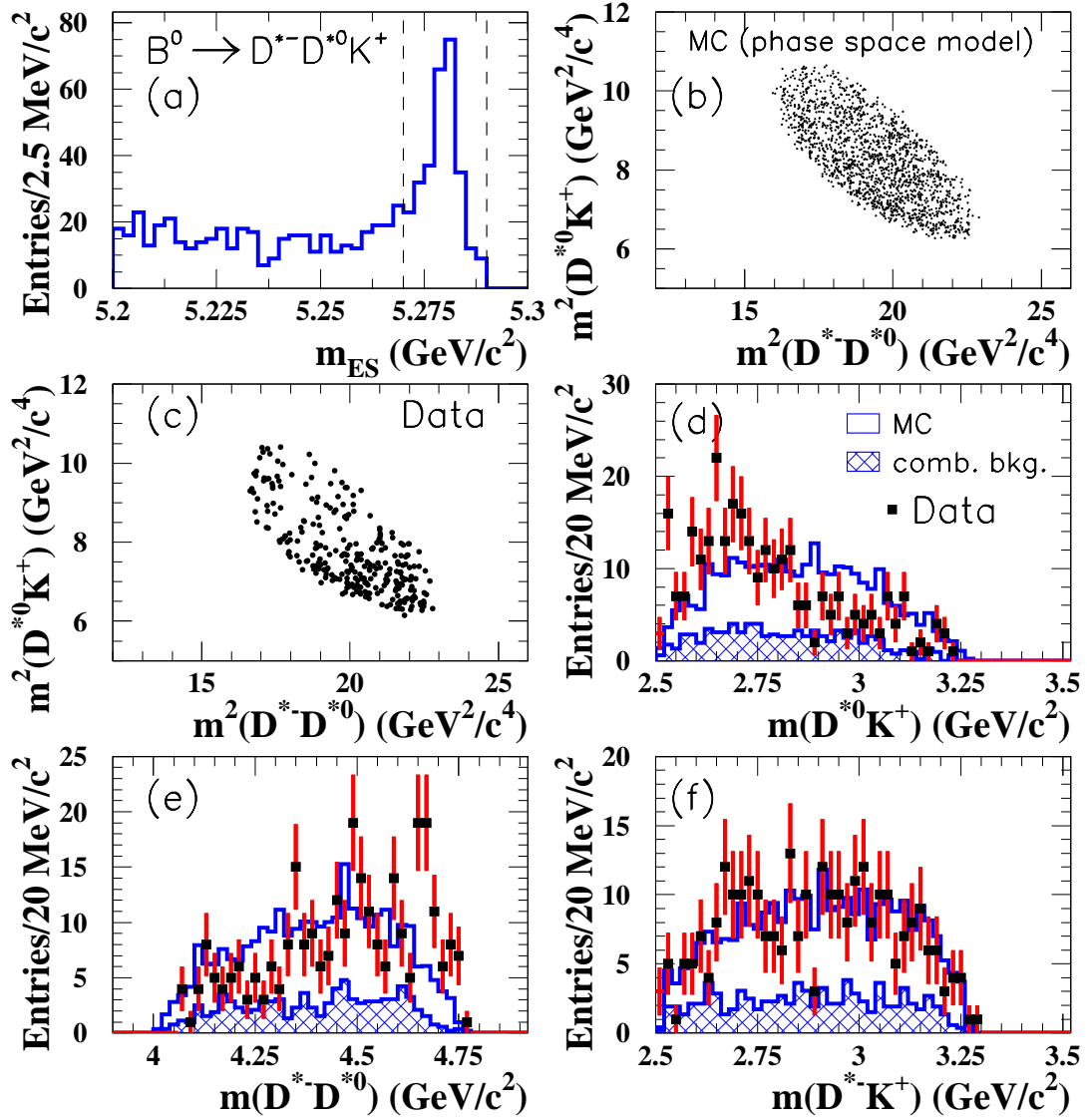,width=0.95\textwidth}
\caption{Dalitz plots and projections for the decay 
$\Bz \to \Dstarm\Dstarz\Kp$. The content of the different plots is discussed in the text. }
\label{Fi:dalitz08}
\end{center}
\end{figure*}
As suggested in Ref.~\cite{ref:colangelo2}, the study of decays \btoddk\  could be 
used to search for evidence of the yet undiscovered broad $j_q = 1/2$ 
$D_{sJ}$ states, if the decays $D_{sJ} \to D^{(*)}K$ are allowed by the 
available phase space. The decay mode $\Bz \to \Dstarm\Dstarz\Kp$, which has the 
largest number of reconstructed events and  also has the largest purity, is 
used for this search.  The results are shown in Fig.~\ref{Fi:dalitz08}.
The upper right plot (Fig.~\ref{Fi:dalitz08} (b)) is the Dalitz plot 
$m^2(\Dstarz\Kp)$ {\it vs.} $m^2(\Dstarm\Dstarz)$ expected for three-body 
$\Bz \to \Dstarm\Dstarz\Kp$ decays generated with a phase 
space model. The Dalitz plot $m^2(\Dstarz\Kp)$ {\it vs.} $m^2(\Dstarm\Dstarz)$ for 
data events in the signal region $5.27<\mes(\Dstarm\Dstarz\Kp)<5.29\gevcc$ 
(Fig.~\ref{Fi:dalitz08} (a))  
is shown in Fig.~\ref{Fi:dalitz08} (c). The next three plots 
(Figs.~\ref{Fi:dalitz08} (d),~\ref{Fi:dalitz08} (e),~\ref{Fi:dalitz08} (f)) show the 
projections $m(\Dstarz\Kp)$, $m(\Dstarm\Kp)$, and 
$m(\Dstarm\Dstarz)$ for the same events. The hatched histograms show the 
contribution expected from the combinatorial background; their shapes are 
derived from the events with $\mes(\Dstarm\Dstarz\Kp)<5.26\gevcc$. 
The open histograms show the contribution expected for three-body 
$\Bz \to \Dstarm\Dstarz\Kp$ decays generated with a phase- 
space model. The density of events in the lower region of the Dalitz plot 
(i.e., for small values of $m(\Dstarz\Kp)$ and large values of 
$m(\Dstarm\Dstarz)$) is significantly larger in the data 
( Fig.~\ref{Fi:dalitz08} (c)) than in the simulation 
with no resonance ( Fig.~\ref{Fi:dalitz08} (b)). 
It could be interpreted as the presence of a broad resonance decaying to 
$\Dstarz\Kp$, like the $J^P=1^+$, $j_q=1/2$ state predicted by Heavy Quark 
Symmetry models \cite{ref:rosner, ref:isgur, ref:godfrey}. However, 
 more events are necessary to confirm this hypothesis and to estimate 
the resonance properties such as mass and width. 

As previously discussed (Sec.~\ref{sec:Systematics}), the hypothetical presence of broad resonances 
in the decay chain is accounted for by a 5\% relative systematic error on all the \btoddk\ branching 
fraction measurements described in this paper.  

\section{Conclusions}
\label{sec:conclusion}

A measurement of the branching fractions for the 22 \btoddk\ modes
is given in Table~\ref{tab:ddkyields}. For the decay modes for which
$S/\sqrt{B}$ is smaller than 4, a 90\% C.L. upper limit
is also derived (here, $B$ is the sum of the combinatorial background and
the cross-feed background from other $\Dbar^{(*)}D^{(*)}K$ modes
and $S=N-B$, where $N$ is the total yield in the signal region).
This is the first complete measurement of all
possible \btoddk\ channels. The measured branching fractions
are in good agreement with earlier measurements made with smaller data sets
for some of these modes \cite{cleoddk,alephddk,babarddk,belleddk}. 

The existence of the decays \modevii\ and \modex, which are an admixture 
of $CP$ even and $CP$ odd eigenstates, has been demonstrated. 
These decay modes could be used in the future, with 
larger event samples, to also determine \stwob
and 
$\cos\! 2 \beta$ \cite{ref:browderddk, ref:colangelo, ref:charles}. 
A significant signal for the color 
suppressed decay mode \modexvi\ has also been observed.

One of the motivations of this analysis is to understand whether decays 
\btoddk\ can explain the 
wrong-sign $D$-meson rates in $B$ decays and reconcile the 
total $ b \to c \cbar s$ rate 
with the predictions of Ref.~\cite{buchalla}. After 
summing over all submodes, the branching 
fractions of the \Bz and of the \Bu to $\Dbar^{(*)} D^{(*)} K$ are found to be
\begin{small}
\begin{equation}
\BR(\Bz\to \Dbar^{(*)} D^{(*)} K) = \left (4.3 \pm 0.3\stat \pm 0.6\syst \right )\%,
\end{equation}
\begin{equation}
\BR(\Bu\to \Dbar^{(*)} D^{(*)} K) = \left (3.5 \pm 0.3\stat \pm 0.5\syst \right )\%.
\end{equation}
\end{small}
This study shows that a significant fraction of the transitions 
$\b \to \c \cbar \s$ proceed 
through the decays $B\to \Dbar^{(*)} D^{(*)}K$. These decay modes account for
about one half of the wrong-sign $D$ production rate in $B$ decays,
$\BR(B \to D\, X)=(7.9\pm 2.2)\% $ \cite{cleoupv}; however, because of
the large statistical error on the latter measurement, it is not yet clear
whether they saturate it. 

A search for resonant substructures shows that the $D_{s1}^+(2536)$ 
contribution to  $B\to \Dbar^{(*)}\Dstarz\Kp$ decays is small. No evidence 
for a $D_{sJ}^+(2573)$ contribution to $B\to \Dbar^{(*)} \Dz\Kp$ decays is 
found. Finally, a simple Dalitz-plot analysis of the decays 
$\Bz \to \Dstarm\Dstarz\Kp$ shows that the three-body phase-space decay 
model does not give a satisfactory description of these decays. 

\section{Acknowledgments}
\label{sec:Acknowledgments}
We are grateful for the 
extraordinary contributions of our \pep2\ colleagues in
achieving the excellent luminosity and machine conditions
that have made this work possible.
The success of this project also relies critically on the 
expertise and dedication of the computing organizations that 
support \babar.
The collaborating institutions wish to thank 
SLAC for its support and the kind hospitality extended to them. 
This work is supported by the
US Department of Energy
and National Science Foundation, the
Natural Sciences and Engineering Research Council (Canada),
Institute of High Energy Physics (China), the
Commissariat \`a l'Energie Atomique and
Institut National de Physique Nucl\'eaire et de Physique des Particules
(France), the
Bundesministerium f\"ur Bildung und Forschung and
Deutsche Forschungsgemeinschaft
(Germany), the
Istituto Nazionale di Fisica Nucleare (Italy),
the Foundation for Fundamental Research on Matter (The Netherlands),
the Research Council of Norway, the
Ministry of Science and Technology of the Russian Federation, and the
Particle Physics and Astronomy Research Council (United Kingdom). 
Individuals have received support from 
the A. P. Sloan Foundation, 
the Research Corporation,
and the Alexander von Humboldt Foundation.

\end{document}